# Title: Old Drugs for JAK-STAT Pathway Inhibition in COVID-19 Patients:

Mohammad Reza Dayer

Department of Biology, Faculty of Science, Shahid Chamran University of Ahvaz, Ahvaz, Iran

Corresponding author:

Mohammad Reza Dayer, Department of Biology, Faculty of Sciences, Shahid Chamran University of Ahvaz, Ahvaz, Iran. Tel/Fax: +98-6113331045, E-mail: mrdayer@scu.ac.ir

**Abstract:**

The pandemic threat of COVID-19 with more than 37 million cases in which about 5 percent entering critical stage characterized by cytokine storm and hyperinflammatory condition, the state more often leads to admission to intensive care unit with rapid mortality. Janus kinase enzymes of Jak-1, Jak-2, Jak-3, and Tyk2 seem to be good targets for inhibition by medications to control cytokine storm in this context. In the present work, the inhibitory properties of different analgesic drugs on these targets are studied to assess their ability for clinical application from different points of view. Our docking results indicated that naproxen, methadone, and amitriptyline considering their higher binding energy, lower energy variance and higher hydrophobicity, seem to express more inhibitory effects on Janus kinase enzymes than thats for approved inhibitors i.e. baricitinib and ruxolitinib. Accordingly, we suggest our wide list of candidate drugs including indomethacin, etodolac, buprenorphine, rofecoxib, duloxetine, valdecoxib, naproxen, methadone, and amitriptilin for clinical assessments for their usefulness in COVID-19 treatment, especially taking into account that up to now, there is no approved cure for this disease.

**Introduction:**

Janus kinase (JAK) is a family of intracellular tyrosine kinase enzymes that participate in signal transduction through cytokine receptors in the JAK-STAT pathway. There are two types of cytokine receptors; type-1, and type-II. Both of these receptors have no kinase activities and so they dependent on JAK enzymes for phosphorylation and signal transduction. The family of JAK is comprised of tyrosine kinase-2 (Tyk2), JAK-1, JAK-2, and JAK-3 enzymes. This enzyme Tyk2 is the first described member of this family. The enzyme collaborates with cytoplasmic domains of cytokine receptors (type I and II) for signal transduction induced by IL-6, IL-11, IFN-α, IFN-β, and IFN-γ cytokines. The JAK-1 enzyme uses the gamma chain of type-I receptor and participates in signal transduction from IL-2, IL-4, IL-7, IL-9, IL-15, and IL-21 cytokines and also mediates signals through type-II receptor posed by IFN-α, IFN-β and IFN-γ.

The JAK-2 enzyme facilitates signal transduction through a type-I receptor that is induced by IL-3, IL-5, IL-6, IL-11, GM-CSF, EPO, TPO, GH, G-CSF and also signals through type-II receptor exerted by TFN-α, IFN-β and IFN-γ cytokines [3-6]. Unlike the other Janus kinase enzymes, JAK-3 only mediates signals through type-I receptor that is induced by IL-2, IL-4, IL-7, IL-9, IL-15, IL-21 cytokines [7-12]. In contemporary medicine, Janus kinase inhibitors are used as medications to interfere with JAK-STAT signaling pathways and to manage and control hyperinflammatory states or cytokine storms in severe diseases such as cancer and autoimmune diseases [13-14]. Among JAK inhibitors some are approved for clinical use including ruxolitinib, against JAK1/JAK2, oclacitinib, against JAK1, baricitinib, against JAK1/JAK2, peficitinib, against JAK3, fedratinib, against JAK2 inhibitor and upadacitinib, against JAK1 pathways [15-18]. There are also some JAK inhibitors, e.g. filgotinib, cerdulatinib, gandotinib, lestaurtinib, momelotinib, pacritinib, and abrocitinib which are in clinical trials for future applications [19-21].

The newly emerging disease of COVID-19 is caused by severe acute respiratory syndrome coronavirus 2 (SARS-CoV-2). The disease causes a pandemic threat with more than 37 million cases and more than 1 million deaths by October 2020 [27-28]. It is well documented that COVID-19 patients experience a dramatic increase in plasma levels of different kinds of inflammatory cytokines that in server cases lead to profound infiltration of immune cells in the lungs with ultimate alveolar damage and death [22-26].

Increasing the cytokines of IL-2, IL-6, IL-7, IL-10, G-CSF, GM-CSF, and IFN-γ in accordance with increasing different chemokines comprises the main cause for COVID-19 mortality, the state primarily mediated by JAK-STAT pathway [28-30].

There are increasing efforts performed to control hyperinflammatory state in COVID-19 by application of Janus kinase inhibitors. Ruxolitinib is one of the approved inhibitors used in the clinic for treatment of myelofibrosis and selectively inhibits JAK-1 and JAK-2 shows reasonable effects on mitigating hyperinflammatory state in COVID-19 patients [27-32]. Baricitinib is the next example of JAK inhibitors that is prescribed as anti-rheumatic drug for rheumatoid arthritis and significantly blocks both JAK1 and JAK2 decreases fever, breathlessness, cough and improves pulmonary function in COVID-19 patients [33-34].

There are also miscellaneous reports indicating the benefits of JAK inhibitors in COVID-19 treatment that encouraged us to search for new candidates among old analgesic or pain relief drugs for their ability in this context from a bioinformatics point of view [35-37].

**Methods and Materials:**

**Coordinate structures for JAK enzymes**:

Coordinate structures of JAK-1, JAK-2, JAK-3, and TyK2 enzymes with PDB IDs' of 4I5C, 2W1I, 3LXK, and 4GVJ, respectively were retrieved from protein data bank (https://www.rcsb.org/). These structures were obtained by the X-ray diffraction and refined at the resolutions of 2.1Å, 2.60Å, 2Å, and 2.03Å, respectively. These structures were energy minimized in separate rectangular boxes with dimensions of 9.79×9.98×6.84nm, 7.87×7.62×9.27nm, 5.62×5.73×6.87nm, and 5.67×7.18×6.18nm dimensions respectively. The boxes were filled with SPCE water. The algorithm of the steepest descent algorithm, neutral pH, 37°C temperature, 1atmosphere of pressure, and total energy of 200kj/mol was used as minimization criteria [38-39].

**Drugs coordinate structures:** The coordinate structures of candidates drugs (selected from analgesics or pain relief drugs including almotriptan, amitriptyline, amlodipine, baricitinib, buprenorphine, celecoxib, diclofenac, duloxetine, ergotamine, esomeprazole, etodolac, famotidine, fentanyl, indomethacin, lansoprazole, lasmiditan, methadone, nalbuphine, naloxone, naproxen, naratriptan, oxycodone, piroxicam, remifentanil, rimegepant, rofecoxib, ruxolitinib, sufentanil, sulindac, tofacitinib, ubrogepant and valdecoxib) in SDF format were retrieved from PubChem database (https://pubchem.ncbi.nlm.nih.gov/) and converted to PDB format with Open Babel software (http://openbabel.org/). The structures then were energy minimized in ArgusLab software (http://www.arguslab.com/) [40].

**Enzymes Active Sites:** The active sites of JAK enzymes were extracted using Computed Atlas of Surface Topography of proteins server (http://sts.bioe.uic.edu/castp/).

**Docking experiments:** In order to study the potential ability of studied drugs in binding to the JAK enzyme active sites we have performed blind docking experiments in Hex 8.0.0 (http://www.loria.fr/~ritchied/hex/) [41]. The setting of sahpe+electrostatic and macro sampling method, optimized structures of JAK enzymes as receptor and drugs as ligands were used for docking experiments. The best 100 docking pose and their binding energies were recorded for analysis.

**Drugs Hydrophobicity:** Partition coefficient or logP is an acceptable index for drug hydrophobicity in which a positive value means the compound has a higher affinity for hydrophobic phase. The server of the Virtual Computational Chemistry Laboratory (http://www.vcclab.org/) was used to calculated logP for the studied drugs [42].

**Data Handling and Analysis**: All the numerical data were exploited in Excel and SPSS software. P-value under 0.05 was considered as the significance level.

**Results and Discussion:**

Figure 1-a represents sequence alignment results for Janus kinase enzymes. As it is clear, there are high sequence similarities between these enzymes in such a way that the overall structures and their binding site motifs indicate structural similarities (Figure 1-b). However, the ligand binding properties of the enzymes are somewhat different that lead to different results seen in our docking experiments.

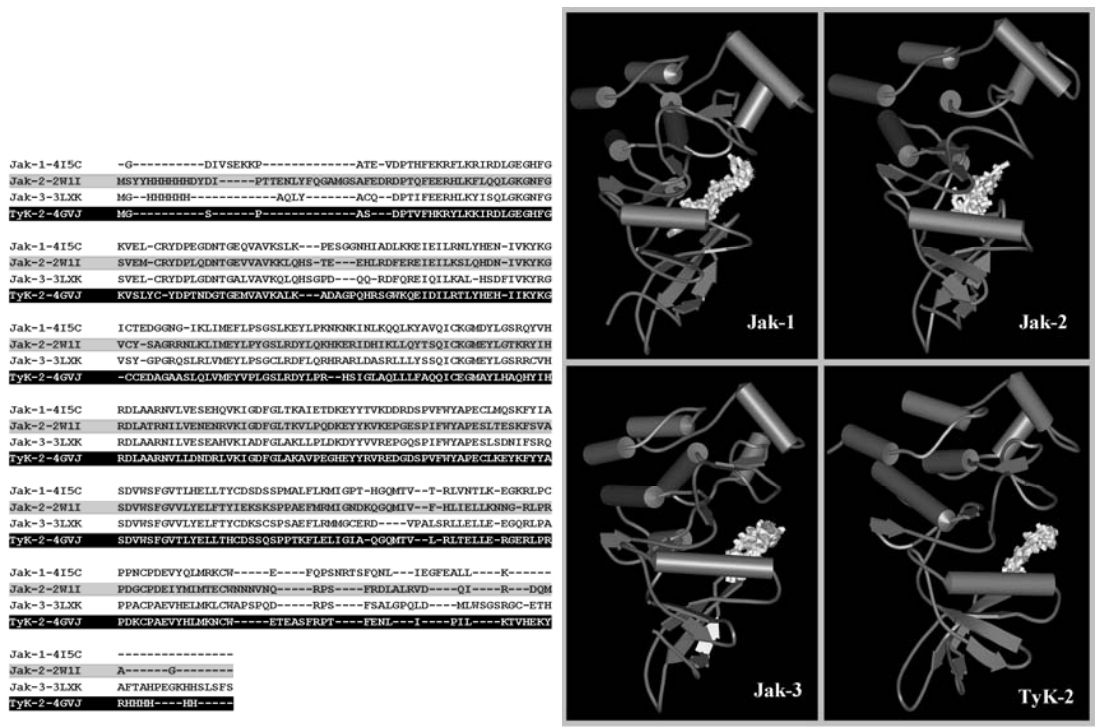

**Figure 1: a-** Multiple sequence alignment for Jak-1, Jak-2, Jak-3, and TyK-2 was performed on CLUSTAL (www.ebi.ac.uk/Tools/msa/clustalo/). **b-** Binding site predicted using Computed Atlas of Surface Topography of proteins server (http://sts.bioe.uic.edu/castp/) for Jak-1, Jak-2, Jak-3, and TyK-2 used for analysis of docking results.

Table 1 represents the docking results obtained for the studied drugs used as ligands and Jak-1, Jak-2, Jak-3, and TyK2 as receptors. The drugs with higher binding energies, lower variances, higher logPs', and higher degree of binding site occupancies seem to be better candidates for enzyme inhibition.

**Table 1: Average binding energy in kJ/mol (as Mean±SD)s well as variance in their binding energy calculated for the best 100 poses for Janus kinase enzyme and the logP values of the drugs calculated on (http://www.vcclab.org/) website.**

|  | JAK-1 | | | JAK-2 | | | JAK-3 | | | TYK2 | | | |
|---|---|---|---|---|---|---|---|---|---|---|---|---|---|
|  | Mean±SD | Variance | Occupancy | Mean±SD | Variance | Occupancy | Mean±SD | Variance | Occupancy | Mean±SD | Variance | Occupancy | logP |
| almotriptan | -344.5±08 | 67.81 | 100 | -299.1±76 | 38.41 | 59 | -323.44±9 | 92.07 | 100 | -322.98±7 | 63.46 | 91 | 2.04 |
| amitriptyline | -312.29±8 | 77.97 | 100 | -286.80±6 | 44.82 | 43 | -296.33±6 | 37.98 | 100 | -288.42±7 | 52.73 | 100 | 5.1 |
| amlodipine | -285.01±13 | 183.16 | 82 | -324.08±8 | 78.87 | 41 | -351.65±9 | 85.57 | 50 | -327.91±11 | 132.51 | 49 | 2.22 |
| baricitinib | -340.86±8 | 74.59 | 100 | -311.13±6 | 41.96 | 24 | -309.04±8 | 74.49 | 97 | -331.52±9 | 90.23 | 100 | 2.22 |
| buprenorphine | -389.48±14 | 206.8 | 100 | -349.98±12 | 150.49 | 21 | -319.16±7 | 51.84 | 68 | -372.33±9 | 90.68 | 98 | 1.08 |
| buprenorphine | -389.48±14 | 206.8 | 100 | -349.69±12 | 153.97 | 18 | -319.16±7 | 51.84 | 58 | -372.33±9 | 90.68 | 100 | 4.53 |
| celecoxib | -538.62±13 | 179.3 | 0 | -444.11±8 | 68.44 | 47 | -530.02±15 | 246.3 | 21 | -426.53±14 | 211.48 | 25 | 4.53 |
| diclofenac | -391.72±9 | 82 | 0 | -328.13±10 | 113.12 | 0 | -392.19±8 | 74.11 | 39 | -304.89±9 | 94.61 | 0 | 4.98 |
| duloxetine | -332.81±11 | 124.64 | 100 | -310.25±8 | 71.65 | 60 | -301.51±9 | 98.96 | 93 | -299.41±8 | 70.77 | 87 | 4.72 |
| ergotamine | -414.33±13 | 178.31 | 0 | -396.76±13 | 186.13 | 85 | -368.09±11 | 122.89 | 80 | -412.71±13 | 189.34 | 93 | 3.99 |
| esomeprazole | -363.94±10 | 103.22 | 100 | -293.53±8 | 70.42 | 28 | -307.21±8 | 74.64 | 94 | -311.32±7 | 55.71 | 74 | 2.95 |
| etodolac | -304.10±8 | 70.5 | 100 | -281.44±8 | 74.01 | 75 | -273.05±7 | 60.78 | 92 | -283.78±8 | 73.04 | 90 | 3.39 |
| famotidine | -299.92±5 | 28.37 | 100 | -264.88±7 | 59.29 | 91 | -266.55±8 | 74.12 | 86 | -274.51±7 | 55.7 | 95 | 1.66 |
| fentanyl | -351.58±6 | 38.54 | 100 | -314.62±8 | 78.45 | 68 | -322.89±10 | 103.79 | 82 | -347.59±8 | 68.84 | 72 | -0.2 |
| indomethacin | -373.64±10 | 113.49 | 99 | -346.31±13 | 178.68 | 80 | -358.89±10 | 119.69 | 16 | -322.96±8 | 71.17 | 96 | 4.25 |
| lansoprazole | -567.43±12 | 146.08 | 100 | -452.15±10 | 100.09 | 25 | -538.29±15 | 243.78 | 46 | -431.28±11 | 135.44 | 11 | 2.84 |
| lasmiditan | -521.15±13 | 177.33 | 77 | -431.78±12 | 149.32 | 4 | -512.18±8 | 76.41 | 72 | -385.99±16 | 267.03 | 40 | 2.76 |
| methadone | -320.14±9 | 99.74 | 100 | -319.62±9 | 82.1 | 86 | -272.80±7 | 52.53 | 95 | -307.68±6 | 43.04 | 91 | 4.14 |
| nalbuphine | -336.72±6 | 46.67 | 100 | -333.39±8 | 66.14 | 79 | -288.05±8 | 67.86 | 63 | -313.93±9 | 95.45 | 57 | 2 |
| naloxone | -319.80±10 | 115.51 | 100 | -318.02±10 | 111.47 | 97 | -286.58±9 | 95 | 95 | -288.36±5 | 32.65 | 80 | 1.47 |
| naproxen | -276.41±7 | 57.86 | 100 | -252.06±6 | 40.54 | 80 | -250.64±6 | 38.96 | 80 | -257.66±5 | 28.89 | 74 | 3.29 |
| naratriptan | -340.01±6 | 41.61 | 100 | -308.16±7 | 50.56 | 30 | -316.33±12 | 162.6 | 94 | -330.94±10 | 108.44 | 100 | 2.16 |
| oxycodone | -301.71±5 | 26.67 | 100 | -287.69±7 | 61.6 | 83 | -274.92±6 | 47.76 | 63 | -281.10±6 | 36.9 | 94 | 1.04 |
| piroxicam | -327.61±7 | 60.33 | 100 | -291.07±8 | 68.25 | 75 | -294.22±11 | 130.14 | 100 | -298.19±6 | 41.28 | 74 | 2.2 |
| remifentanil | -368.40±11 | 136.09 | 100 | -322.15±7 | 49.44 | 70 | -305.59±8 | 64.79 | 94 | -335.01±9 | 89.36 | 94 | 1.75 |
| rimegepant | -513.18±18 | 329.07 | 97 | -437.91±12 | 168.09 | 24 | -462.48±10 | 117.64 | 0 | -448.28±10 | 104.52 | 100 | 2.68 |
| rofecoxib | -301.84±5 | 29.01 | 100 | -275.29±6 | 38.81 | 58 | -274.34±8 | 69.14 | 89 | -285.52±5 | 34.76 | 95 | 2.32 |
| ruxolitinib | -331.57±8 | 79.9 | 100 | -302.82±12 | 152.08 | 0 | -287.14±8 | 77.19 | 0 | -309.28±7 | 58.25 | 95 | 2.94 |
| sufentanil | -364.00±8 | 78.92 | 100 | -336.96±11 | 133.74 | 39 | -330.00±10 | 102.77 | 95 | -350.91±12 | 146.58 | 100 | 3.4 |
| sulindac | -373.87±8 | 75.99 | 10 | -332.61±10 | 105.94 | 0 | -362.92±6 | 42.18 | 100 | -315.03±7 | 56.63 | 67 | 2.96 |
| tofacitinib | -336.07±9 | 94.76 | 100 | -300.26±5 | 34.57 | 41 | -304.10±11 | 131.89 | 89 | -307.66±6 | 39.09 | 92 | 1.58 |
| ubrogepant | -609.70±24 | 615.54 | 0 | -509.55±14 | 216.56 | 60 | -564.97±17 | 300.73 | 44 | -472.12±23 | 574.14 | 4 | 3 |
| valdecoxib | -299.65±7 | 60.1 | 100 | -275.96±6 | 39.86 | 40 | -281.01±5 | 34.15 | 98 | -288.52±7 | 55.06 | 95 | 3.32 |

In order to optimize the calculated variables in table 1 and to obtain a reliable and cumulative index for comparing the studied drugs from their inhibitory potency point of view, we converted the values of binding energies, 1/variances (as stability index), percent of binding site occupancies and the logP values to normalize absolute values in 0 to 1 range and then summate them in a total cumulative index for this purpose (Table 2). It is of prime importance to note that, in these calculations we take the same contribution effect for all variables on the total effects of drugs, the assumption that could be under question quantitatively.

Table 2: The total cumulative index for each drug calculated based on normalized values for binding energies, 1/variances, binding site occupancies, and logP as described in the text.

|  | Jak-1 | Jak-2 | Jak-3 | Tyk2 | Total |
|---|---|---|---|---|---|
| fentanyl | 2.23 | 1.72 | 1.68 | 1.84 | 7.47 |
| amlodipine | 1.87 | 1.93 | 1.96 | 1.84 | 7.60 |
| buprenorphine | 1.98 | 1.34 | 2.12 | 2.30 | 7.74 |
| ubrogepant | 1.63 | 2.37 | 2.14 | 1.68 | 7.82 |
| ruxolitinib | 2.45 | 1.40 | 1.53 | 2.68 | 8.06 |
| diclofenac | 1.94 | 1.93 | 2.52 | 1.93 | 8.32 |
| lasmiditan | 2.32 | 1.66 | 2.61 | 1.87 | 8.46 |
| lansoprazole | 2.67 | 2.05 | 2.11 | 1.79 | 8.62 |
| rimegepant | 2.42 | 1.84 | 1.63 | 2.75 | 8.64 |
| sulindac | 1.64 | 1.56 | 3.03 | 2.43 | 8.66 |
| nalbuphine | 2.52 | 2.38 | 2.04 | 1.93 | 8.86 |
| naloxone | 2.04 | 2.22 | 2.10 | 2.58 | 8.96 |
| celecoxib | 1.92 | 2.75 | 2.18 | 2.18 | 9.02 |
| tofacitinib | 2.14 | 2.32 | 2.00 | 2.62 | 9.08 |
| naratriptan | 2.62 | 2.02 | 2.13 | 2.39 | 9.17 |
| remifentanil | 2.14 | 2.40 | 2.35 | 2.32 | 9.21 |
| baricitinib | 2.35 | 2.12 | 2.41 | 2.46 | 9.34 |
| esomeprazole | 2.43 | 1.93 | 2.52 | 2.50 | 9.38 |
| piroxicam | 2.41 | 2.28 | 2.21 | 2.50 | 9.41 |
| oxycodone | 2.70 | 2.19 | 2.04 | 2.52 | 9.44 |
| ergotamine | 1.61 | 2.62 | 2.51 | 2.74 | 9.49 |
| famotidine | 2.76 | 2.37 | 2.12 | 2.38 | 9.62 |
| almotriptan | 2.36 | 2.50 | 2.34 | 2.45 | 9.65 |
| sufentanil | 2.60 | 1.99 | 2.53 | 2.61 | 9.73 |
| indomethacin | 2.67 | 2.53 | 1.91 | 2.88 | 10.00 |
| etodolac | 2.54 | 2.46 | 2.63 | 2.56 | 10.19 |
| buprenorphine | 2.66 | 1.98 | 2.69 | 3.00 | 10.33 |
| rofecoxib | 2.87 | 2.48 | 2.32 | 2.84 | 10.52 |
| duloxetine | 2.69 | 2.64 | 2.73 | 2.84 | 10.89 |
| valdecoxib | 2.59 | 2.47 | 3.13 | 2.74 | 10.92 |
| **naproxen** | **2.56** | **2.82** | **2.77** | **2.93** | **11.07** |
| **methadone** | **2.60** | **2.75** | **2.89** | **3.04** | **11.29** |
| **amitriptyline** | **2.85** | **2.78** | **3.42** | **3.16** | **12.21** |

**Conclusion:**

Based on our finding and as indicated in table 2, indomethacin, etodolac, buprenorphine, rofecoxib, duloxetine, valdecoxib, naproxen, methadone, and amitriptyline with higher total effects seem to be good candidates for further studies in JAK-STAT pathway blockage and cytokine storm control in chronic and severe disease of cancer, autoimmune and COVID-19 disease via clinical trials assessments.

**Acknowledgements**

The author would like to express his thanks to the vice chancellor of research and technology of Shahid Chamran University of Ahvaz for providing financial support of this study under Research Grant No: SCU.SB98.477.

[9] Mishra J, Verma RK, Alpini G, Meng F, Kumar N (November 2013). "Role of Janus kinase 3 in mucosal differentiation and predisposition to colitis". The Journal of Biological Chemistry. 288 (44): 31795–806. doi:10.1074/jbc.M113.504126. PMC 3814773. PMID 24045942.

[10] Mishra J, Kumar N (June 2014). "Adapter protein Shc regulates Janus kinase 3 phosphorylation". The Journal of Biological Chemistry. 289 (23): 15951–6. doi:10.1074/jbc.C113.527523. PMC 4047368. PMID 24795043.

[11] Mishra J, Verma RK, Alpini G, Meng F, Kumar N (December 2015). "Role of Janus Kinase 3 in Predisposition to Obesity-associated Metabolic Syndrome". The Journal of Biological Chemistry. 290 (49): 29301–12. doi:10.1074/jbc.M115.670331. PMC 4705936. PMID 26451047.

[12] Mishra, Jayshree; Das, Jugal Kishore; Kumar, Narendra (2017). "Janus kinase 3 regulates adherens junctions and epithelial mesenchymal transition through β-catenin". Journal of Biological Chemistry. 292 (40): 16406–16419. doi:10.1074/jbc.M117.811802. PMC 5633104. PMID 28821617.

[13] Kontzias A, Kotlyar A, Laurence A, Changelian P, O'Shea JJ (August 2012). "Jakinibs: a new class of kinase inhibitors in cancer and autoimmune disease". Current Opinion in Pharmacology. 12 (4): 464–70. doi:10.1016/j.coph.2012.06.008. PMC 3419278. PMID 22819198.

[14] Norman P (August 2014). "Selective JAK inhibitors in development for rheumatoid arthritis". Expert Opinion on Investigational Drugs. 23 (8): 1067–77. doi:10.1517/13543784.2014.918604. PMID 24818516.

[15] Vaddi K, Sarlis NJ, Gupta V (November 2012). "Ruxolitinib, an oral JAK1 and JAK2 inhibitor, in myelofibrosis". Expert Opinion on Pharmacotherapy. 13 (16): 2397–407. doi:10.1517/14656566.2012.732998. PMID 23051187.

[16] Zerbini CA, Lomonte AB (May 2012). "Tofacitinib for the treatment of rheumatoid arthritis". Expert Review of Clinical Immunology. 8 (4): 319–31. doi:10.1586/eci.12.19. PMID 22607178.

[17] Gonzales AJ, Bowman JW, Fici GJ, Zhang M, Mann DW, Mitton-Fry M (August 2014). "Oclacitinib (APOQUEL(®)) is a novel Janus kinase inhibitor with activity against cytokines involved in allergy". Journal of Veterinary Pharmacology and Therapeutics. 37 (4): 317–24. doi:10.1111/jvp.12101. PMC 4265276. PMID 24495176.
10